\documentclass{egpubl}
\usepackage{eurovis2026}
\usepackage{subcaption}
\EuroVisShort  

\usepackage[T1]{fontenc}
\usepackage{dfadobe}

\usepackage{cite}  
\BibtexOrBiblatex
\electronicVersion
\PrintedOrElectronic
\ifpdf \usepackage[pdftex]{graphicx} \pdfcompresslevel=9
\else \usepackage[dvips]{graphicx} \fi

\usepackage{egweblnk}

\title[The Worst Weather In America]%
      {``The Worst Weather In America'': Augmenting the Information Design of Extreme Cold Weather Forecasts}

\author[Correll et al.]
{\parbox{\textwidth}{\centering Michael Correll$^{1}$\orcid{0000-0001-7902-3907},
        Jay Broccolo$^{2}$\orcid{0009-0007-8141-7535},
        Drew Bush$^{2}$\orcid{0000-0002-2826-9445}
        }
        \\
{\parbox{\textwidth}{\centering $^1$ Northeastern University\\
         $^2$ Mount Washington Observatory
       }
}
}

\begin{document}


\maketitle
\begin{abstract}
   Mount Washington is home to extreme, and extremely volatile, weather conditions. Consulting a weather forecast of conditions at the summit is vital for making one's visit as safe as possible. Using the discussion and suggestions arising from a participatory workshop as input, we test a design intervention employing color-coded hazard icons to function as visual summaries of Mount Washington Observatory's current text-heavy forecast through a crowd-sourced study. We find that the use of icons increases the perceived risk of activities involving visiting the mountain. However, we highlight remaining questions around visualization design and design ethics that warrant further study in the domain of how best to communicate cold weather hazards in ways that are mindful of the diversity of literacies and experiences of visitors.

\ccsdesc[500]{Human-centered computing~Empirical studies in visualization}
\ccsdesc[500]{Human-centered computing~Information visualization}

\printccsdesc   
\end{abstract}  
\section{Introduction}

\begin{figure}
    \centering
    \begin{subfigure}{0.6\columnwidth}
        \includegraphics[width=0.9\linewidth]{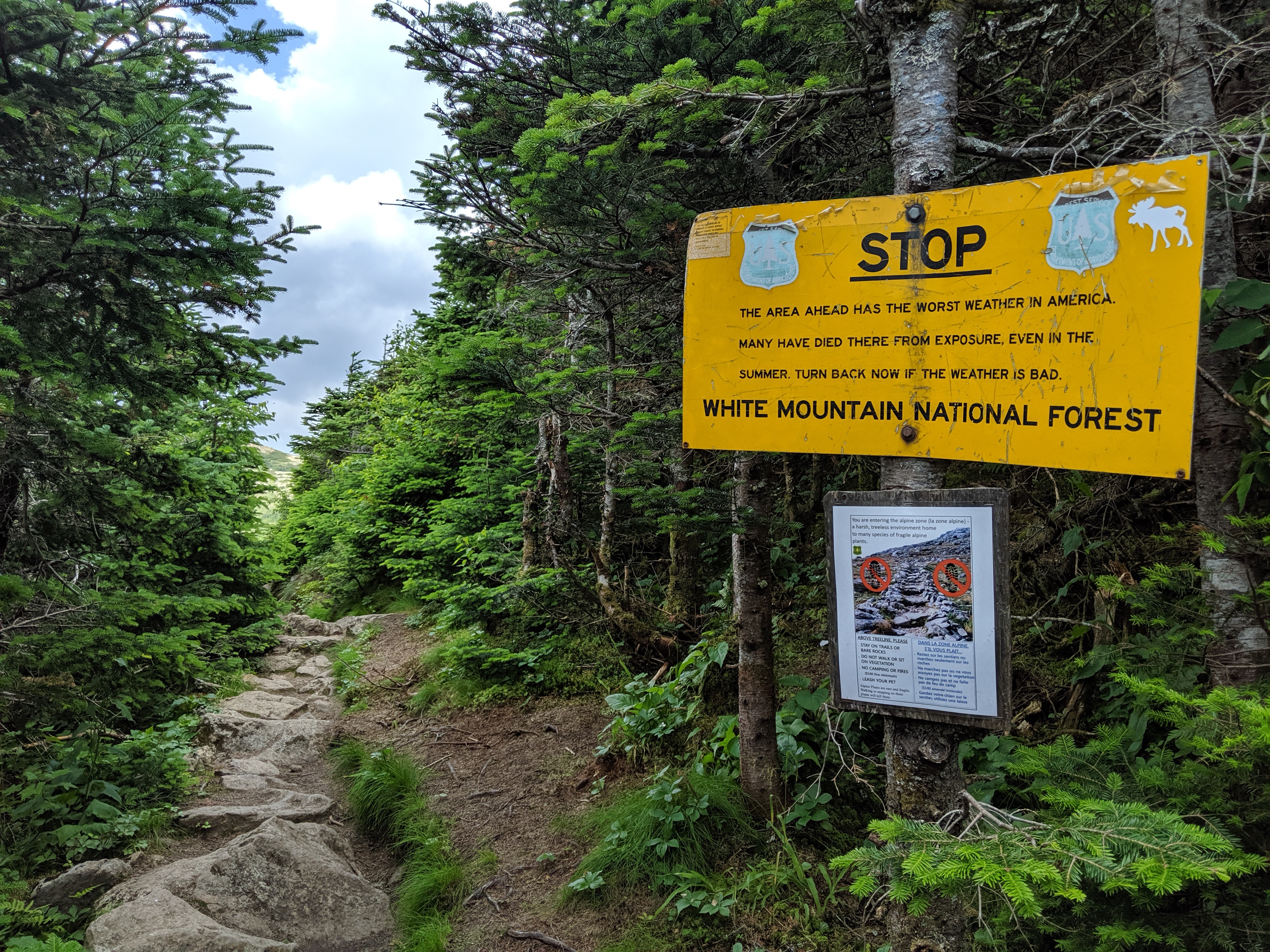}
        \caption{}
        \label{fig:sign}
    \end{subfigure}
    \begin{subfigure}{0.34\columnwidth}
        \includegraphics[width=0.9\linewidth]{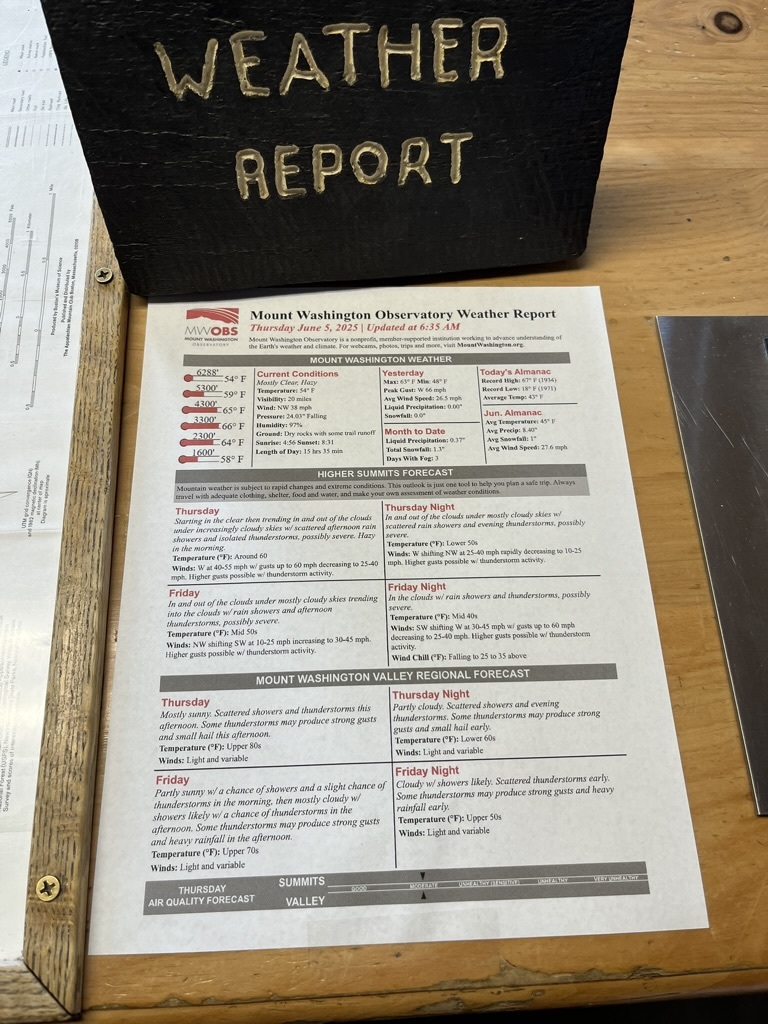}
        \caption{}
        \label{fig:paper}
    \end{subfigure}
    \caption{\ref{fig:sign}: Warning sign upon entering the alpine zone of the Presidential Range. Photo from Wikimedia Commons user ``Sayden'', CC BY-SA 4.0 license\protect\cite{dangersign}.\\
    \ref{fig:paper}: Daily fliers at Mount Washington Observatory showing current conditions and the Higher Summits Forecast. Photo by authors.}
    \label{fig:photos}
\end{figure}

Mount Washington in the U.S. Presidential Range has been listed as one of the top ten deadliest mountains in the world with 200 yearly rescues of hikers in distress and an average of over one death a year in 176 years of records\cite{wiki:Mount_Washington}. One contributing factor to the dangers of the mountain is its extreme and volatile weather, including some of the highest winds ever recorded, giving rise to the title of ``home of the world's worst weather''~\cite{worstweather} (\autoref{fig:sign}). Effectively communicating predictions of hazardous weather conditions is a life-and-death matter. Currently, weather forecasts for the next 48 hours (the high variability reduces the utility of longer-term forecasts) are presented in the Higher Summits Forecast~\cite{highersummits}, updated twice a day by the staff of the Mount Washington Observatory (MWOBS) and available in a variety of formats, including online, printed out by the summit hiker check-in station (\autoref{fig:paper}), text message, and streaming audio. But, despite the information provided by these forecasts, visitors to the mountain continue to make risky decisions that put lives in danger. 

In this paper, we present the results of both a participatory workshop and a crowd-sourced study to identify current challenges and areas of improvement for the MWOBS Higher Summits Forecast. We find that, despite high variability and individual differences in assessing risk from cold weather conditions, augmenting text forecast data with visual overviews can better highlight weather hazards and risks. However, we identify challenges in balancing the varied needs of audiences, their weather and graphical literacies, the information available to meteorologists, and the ontological uncertainty of attempting to visualize the safety of what are inherently risky activities. Most pressingly, we suggest that in design settings like these, where human life is at stake and there are both descriptive and rhetorical goals to be met, designers should opt for limited and conservative rather than ``heroic''~\cite{correll2020we} interventions.

\section{Participatory Workshop}
\begin{figure}
    \centering
    \includegraphics[width=0.95\columnwidth]{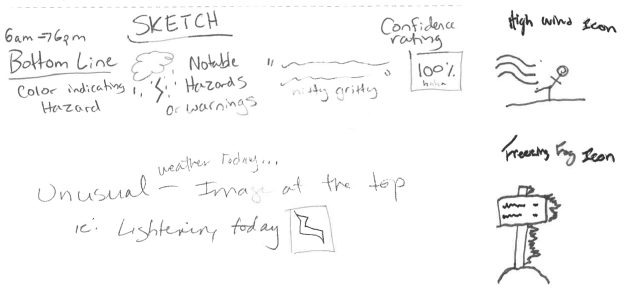}
    \caption{Sketches from workshop participants of proposed visual icons. Icons were proposed both to help summarize forecast information but also to highlight or contextualize hazards specific to Mount Washington that may be unfamiliar to visitors.}
    \label{fig:icons}
\end{figure}

We conducted a hybrid workshop in the Mount Washington area, recruiting from a pool of local organizational users of MWOBS forecasts, including representatives from government agencies like New Hampshire Fish \& Game, the United States Forest Service, research groups, land conservation organizations, hospitality industry, and recreational organizations. 17 people attended the one-day workshop in person, with 5 attending virtually. After an introduction and a brief lecture on uncertainty and weather communication, participants were divided into 5 groups and asked to sketch and brainstorm around a series of short tasks, such as describing the audience of the Higher Summits forecast, their ``blue sky'' ideal forecast formats, a ``red team'' exercise of potential failures, sketching, and a reporting and discussion period with other groups. Workshop materials including facilitator slides and participant sketches are included as supplemental materials.

An initial issue was the length and wordiness of the current forecast. ``Too much text'' was the first piece of feedback. Other challenges include the sheer variability and uncertainty of conditions on the mountain, which preclude usefully accurate long range forecast of the kinds that visitors might be used to in other weather websites or phone apps. Per one participant, `` `Everything can happen' is not a good forecast.'' Connected to this challenge is that the extreme nature of the weather on Mount Washington is outside of most people's common experience. To use an example brought up during the workshop, while a visitor might understand that 70MPH (112KPH) winds with gusts in the triple digits (a not uncommon event in our corpus of forecasts) are high, they might not recognize that, above the treeline with no cover, and in conjunction with low temperatures, even walking upright is difficult if not impossible, and frostbite can set in within five minutes on exposed skin. Conceptualizing such extreme events may involve what Hullman et al.~\cite{hullman2018improving} call ``concrete re-expressions'' of the data values in more familiar terms, or the ``visceralization''~\cite{lee2020data} of data values experientially (for instance, a participant suggested linking to short videos of people attempting to walk in various wind conditions to drive home how dangerous summit winds can be).

A recurring suggestion was the use of visual summaries or other (in the words of one participant) ``filters for opting out of nuance.'' \autoref{fig:icons} shows some participant sketches of proposed icons for depicting hazards or otherwise augmenting the forecast with additional information such as forecast confidence or an overall danger ranking. Proposed sources of hazard and risk information varied, from the judgment of individual forecasters, to AI-based summarizations, to the use of rule-based warnings (e.g., one participant suggested a triad of factors in wind, visibility, and temperature. If one factor is at a dangerous level, then a visitor should take caution. If two or more are at dangerous levels, then the situation might be a ``no go.''). Despite the call for some sort of summarization, it was a matter of contention among participants the extent to which such warnings should be \textit{descriptive} versus \textit{prescriptive}. In the words of one participant, when it comes to summarizing and interpreting forecast data, ``translation is helpful, but opinion is not.'' 

While some participants brainstormed a variety of more dramatic potential re-designs of the forecast, including 3D models and the incorporation of social vetting information (e.g., hikers could leave Yelp-style reviews of current trail conditions), there were competing pressures to be \textit{conservative} in forecast redesigns, in order to preserve the utility of the current forecast, and to avoid going too far into opinion or advocacy, and to ease the burden on the relatively small MWOBS team. A repeated concern was the high stakes, and so large ethical burden, involved in forecast communication where risks to life and limb were a distinct possibility. Echoing language used in other work on visualization ethics~\cite{schwabish2021no}, there was a call to ``first, do no harm''--- to preserve the hard-won trust and credibility of the forecasts put out by the MWOBS team.

\section{Proposed Intervention}

Given the recurring feedback of the need for more rapid summarization and overview of forecast data, as well as the repeated suggestion of icon-based hazard overviews to alert visitors to dangers that might otherwise be buried in the forecast text, we chose to experiment with augmenting the current Higher Summits forecast with a set of color-coded hazard icons. There is a great deal of literature around the communication of uncertainty and risk in weather forecasts; see Ripberger et al. for a recent survey~\cite{ripberger2022communicating}. Icons specifically have centuries of precedence in communicating and summarizing weather conditions for mass audiences~\cite{keeling2010visualization,monmonier1988telegraphy}. We note, however, that icon-based forecasts are one area where prior lit suggests potential challenges in communicating uncertainty, such as the large variability in mapping weather icons to actual numerical probabilities~\cite{reed2022weather}, and the tendency of forecasts making heavy use of icons to flatten and abstract away nuance. Per Zabini~\cite{zabini2016mobile}, quoting Fischoff~\cite{fischhoff1994forecasts}, such forecasts can be guilty of ```immodesty' (`not admitting that sometimes predictions may fail') and `impoverishment' (`not addressing the broader context in which forecasts (and following decisions) are made')''. We also echo the plea in the visualization community to ``give text a chance''~\cite{stokes2024give}, and, rather than making the current text forecast subordinate to a visual display, to consider how visual and textual elements can function in harmony~\cite{stokes2025analysis}.

In addition to these concerns over possible misinterpretation, we had more pragmatic concerns around not overburdening the current MWOBS team, augmenting (rather than replacing) the current forecast narratives, and avoiding being too prescriptive or variable in communication. We therefore focused on several constraints:
\begin{enumerate}
    \item The icons should not involve individual judgment, but have \textbf{consistent meanings}, ideally tied to national or international standards. That is, where possible, rather than forcing individual forecasters to make qualitative judgments of risk, icons should be driven by quantitative scales in use by weather services.
    \item Since collecting new data is in some cases impossible, and in other cases would add additional burdens to a small team with an already difficult job, the icons should be derived only from information \textbf{within the forecast}. Additional hazards, such as areas with avalanche warnings, are important, but are not derivable from the weather forecast \textit{per se}. In many cases, other resources exist to supplement the Higher Summits Forecast (such as the Mount Washington Avalanche Center, which produces a separate avalanche hazard forecast~\cite{MWAC}).
\end{enumerate}

\begin{figure}
    \centering
    \includegraphics[width=0.5\columnwidth]{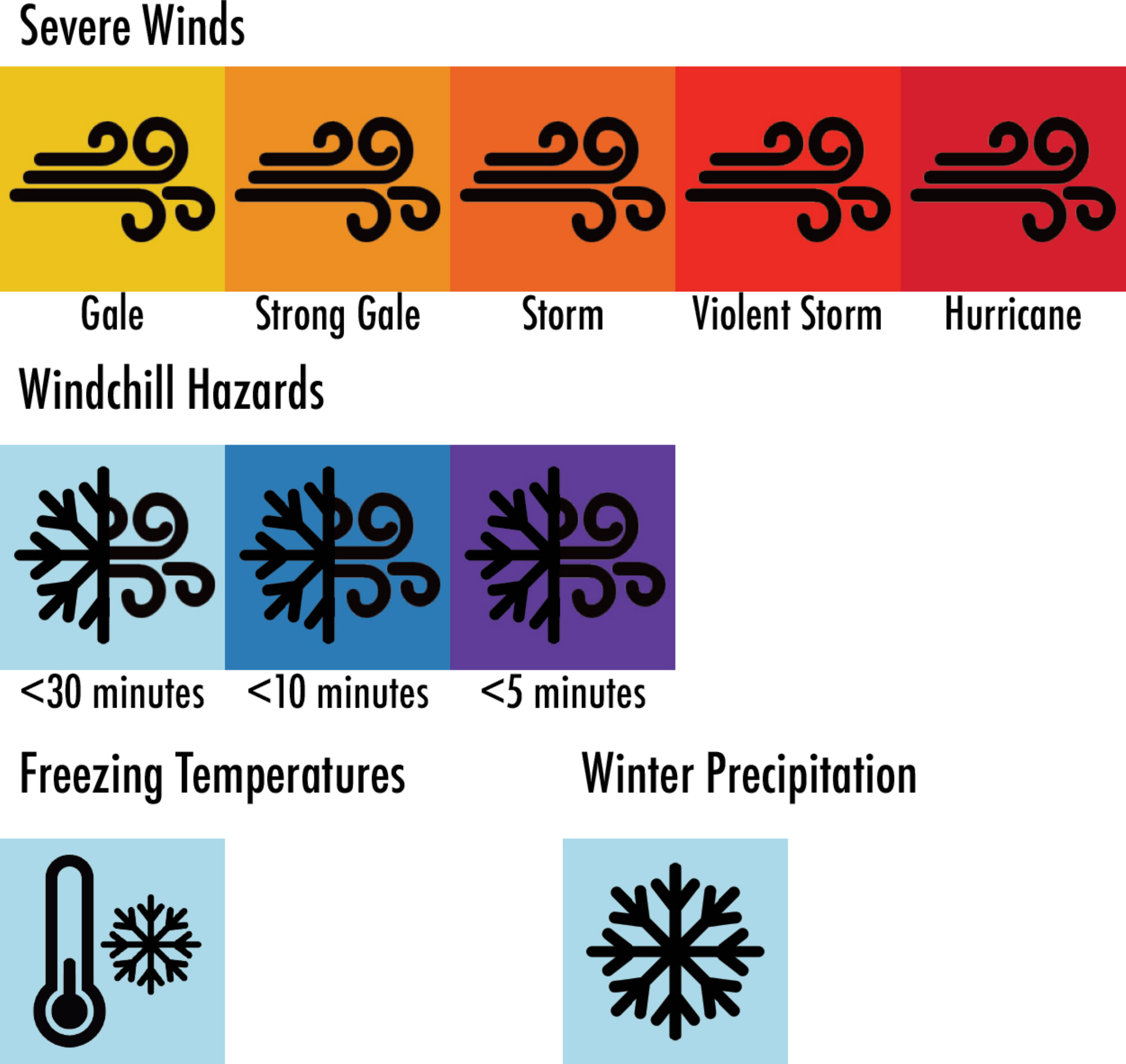}
    \caption{The palette of icons used for our graphical conditions. The wind icon levels and colors use the Beaufort Scale\protect\cite{wiki:Beaufort_scale}. The wind chill levels and colors use the hazard chart produced by the U.S. National Weather Service\protect\cite{nws-windchill}. Freezing temperatures and winter precipitation include any predicted temperature below freezing and any prediction of snow, sleet, or freezing rain during the forecast period, respectively. Note that there are additional hazards (such as avalanche risk) which are not quantified in our forecast datasets, but are known sources of hazards.}
    \label{fig:icon-palette}
\end{figure}

These constraints led us to a limited set of hazard icons (\autoref{fig:icon-palette}), focusing on information that can be derived entirely from predicted wind speed, temperatures, and precipitation, and where existing hazard scales are in wide currency. Our workshop participants had a wide ``weather diet'', consulting multiple sources in addition to the High Summits Forecast. As such, and given the limited bandwidth and wide range of expertise of our intended users, we did not want to introduce new visual conventions. Each icon therefore features an image in the same simplified style as in use by the existing MWOBS products, which, in turn, are based on a standard repository of icons from the Noun Project (\url{https://thenounproject.com/}). The colors in the background of each icon, in turn, are taken from existing national weather hazard guides or weather scales. To avoid the risk of ``double barreled'' icons~\cite{reed2022weather} that try to communicate more than one element of a weather condition simultaneously (e.g., icons which attempt to communicate both the likelihood of snow as well as the amount of snow expected to fall), each hazard is mapped to exactly one icon, and multiple hazards are denoted by multiple icons in sequence. 


\section{Crowd-Sourced Study}

\begin{figure}
    \centering
    \includegraphics[width=\columnwidth]{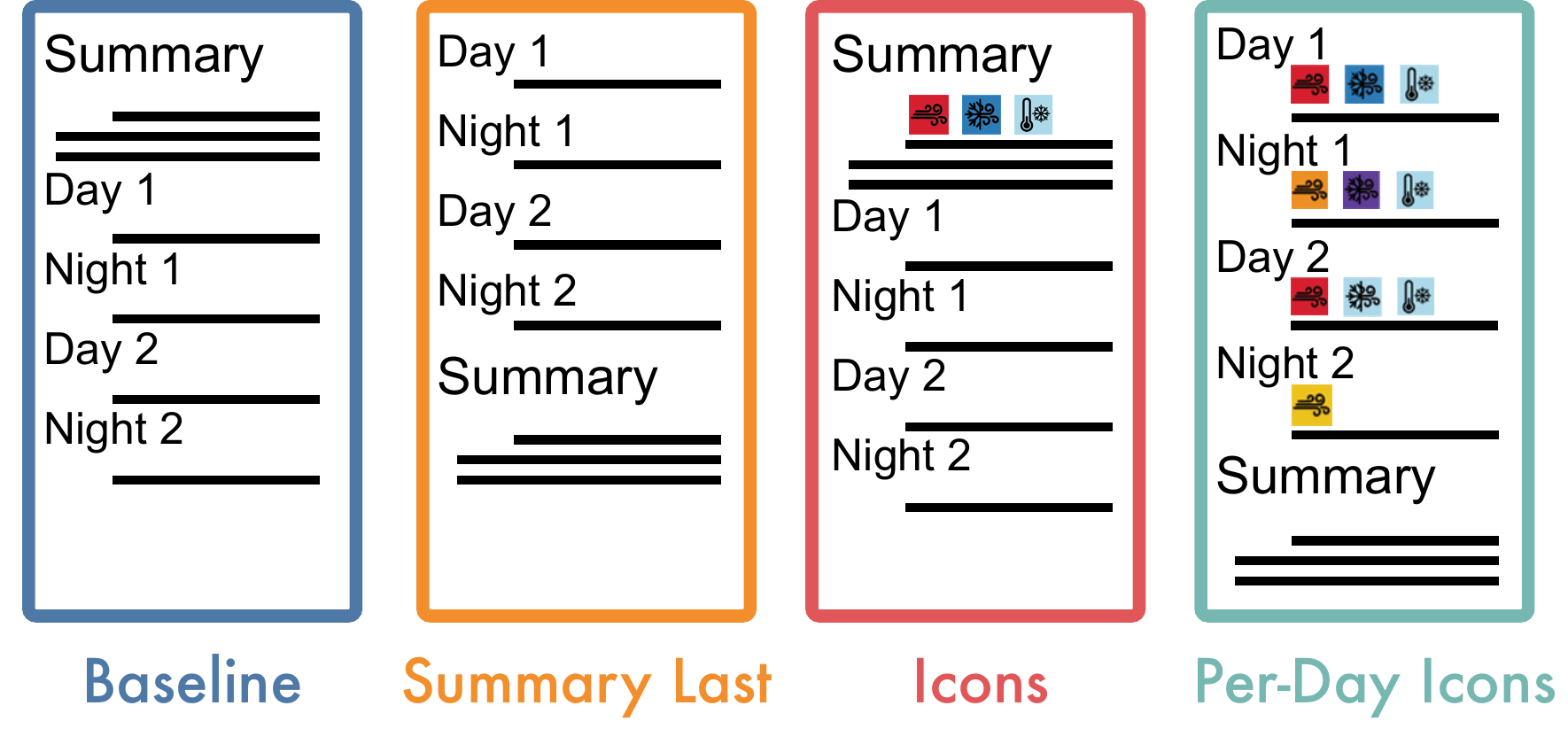}
    \caption{The four forecast design conditions of our experiment. \textbf{Baseline} is the current design, with a textual summary of upcoming conditions followed by shorter predictions of winds, temperature, and wind chill for the next two days and nights. 
    \textbf{Summary Last} reverses the order, presenting the day and night information first; in addition to representing a specific suggestion from the workshop participants, this presentation order more closely models the common admonition of ``overview first, zoom and filter, and then details on demand''~\cite{shneiderman2003eyes}.
    \textbf{Icons} shows hazard icons for the worst conditions predicted across the entire 48-hour interval.
    \textbf{Per-Day Icons} combines the prior two conditions, presenting day and night information first, but with hazard icons for each 12 hour period.}
    \label{fig:conditions}
\end{figure}

We conducted a crowd-sourced study on the affect of our hazard icons on risk assessment in forecasts. Additional data, including all survey materials, stimuli, and anonymized participant data, are included in the supplemental material. We chose a set of five representative forecasts (more details on forecast selection in the supplement) containing hazards like dangerous wind chill, freezing temperatures, snow, flooding, and high winds and gusts. Participants saw forecasts for all five days in random order. The design of the forecast was a between-subjects condition, exploring both the order of forecast information (whether the longer narrative forecast summary was presented first, or if the short day and night predictions were presented first), and the presence or absence of colored icons to function as graphical summaries. \autoref{fig:conditions} shows mockups of these conditions. To better contextualize the ratings from our crowdsourced participants, we also gave the baseline version of our survey to a group of five Mount Washington weather observers stationed on the summit weather station. These results are the ``Observers'' condition; given the low sample size, we exclude their data from our quantitative analyses. Excluding observers, we recruited a total of 128 participants from the Prolific crowdworking platform (63 men, 60 women, and 5 with other gender identities, $M_{age}=42$, $SD_{age}=11.1$).

For each forecast, participants were asked ``Assuming that you are an experienced outdoorsman, rate the relative risk or danger of the following activities, given the forecast information above'' for a list of six activities common to Mount Washington: car trips to the summit, day hikes, mountaineering, backcountry skiing, and single- and multi-night camping trips. We focus in the main paper on aggregate risk for all activities, which, since each activity was assessed on a 0-100 scale, results in responses from 0-600, with 600 being a day where all activities are perceived as maximally risky. In addition to quantitative metrics, we also solicited per-day free text responses: ``What information in the forecast contributed to your decisions?'' and ``What information would you still need in order to be more certain in your decisions?''.
At the end of the survey, in addition to demographics information and free text comments, we a GRiPS risk propensity scale measure~\cite{zhang2019development}. Higher risk propensity was correlated with lower aggregate risk assessments, but this relationship was marginal ($R^2=0.03$, $p=0.053$).

\begin{figure}
    \centering
    \includegraphics[width=\columnwidth]{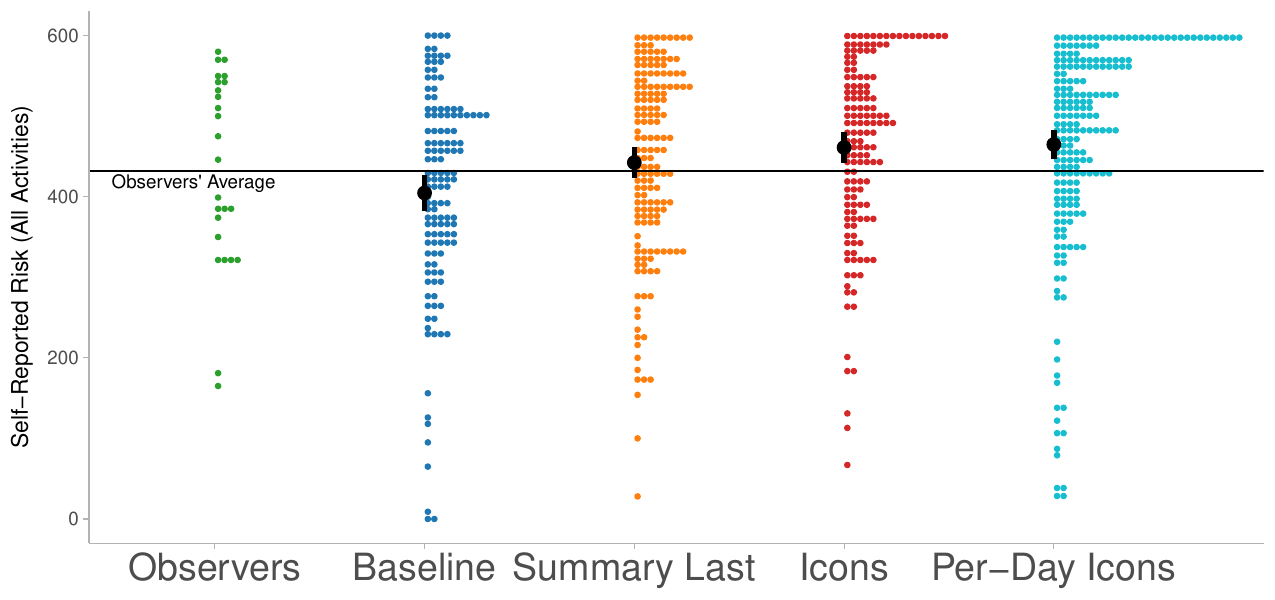}
    \caption{Perceived risk across all activities across our four conditions. Error bars are 95\% t-confidence intervals of the mean. The use of icons resulted in significantly higher risk ratings. The black line indicates the average assessed risk of the same data as provided by five members of the MWOBS forecasting team (green dots).}
    \label{fig:results}
\end{figure}

We performed a one-way repeated measures ANOVA and found that forecast design had a significant impact on perceived risk ($F(3,124)=3.151$,$p=0.027$). A post-hoc set of paired t-tests with a Bonferroni correction applied found that participants in the \textbf{Baseline} text-only condition assessed mountain activities as significantly less risky ($M=404$, $SD=130$) than participants in the \textbf{Icons} condition ($M=461$, $SD=114$, $p=0.001$) and the \textit{Per-Day Icons} condition ($M=465$, $SD=131$, $p=0.0001$), although this effect was marginal compared to those in the \textbf{Summary Last} text-only condition ($M=442$, $SD=120$, $p=0.06$). \autoref{fig:results} shows these results in detail.

A danger brought up during the workshop was that emphasizing per-day summaries would lead to visitors ignoring the longer text forecast (a participant called it the ``tl;dr problem''). As such, the first author coded the free text responses for each participant to assess whether or not the participant mentioned information that could be gleaned from the per-day information and icons alone (such as temperature, wind chill, and wind) and also if the participant mentioned information that was included exclusively in the longer forecast text (such as flood warnings, visibility and fog information, etc.). Across all conditions, nearly all participants (124/128 = 96.88\%) specifically mentioned an element from the per-day information, usually wind, temperature, and/or precipitation when justifying their risk assessments. Participants were less likely to mention information from the longer summary (67/128 = 52.34\%), but, contrary to expectation, and with an acknowledgment that this measure is an undercount (since wind, temperature, and precipitation were discussed in both the overall and per-day summaries, but this code captures only information \textit{exclusively} in the longer summary), this lack of inclusion of information from the detailed forecast was not simply linked to the presence or absence of visual summaries: the \textit{Icons} condition the highest rate of reported use of detailed forecast information (17/29 = 58.62\%) and the \textit{Per-Day Icons} condition the lowest (19/41 = 43.64\%). We also note that over a quarter of participants (36/128 = 28.35\%), in their free text responses, mentioned the wordiness of the forecasts.

As a last comment on our results, we note the presence of both high variability and outliers in our risk assessments as seen in \autoref{fig:results}. Self-assessed risk is inherently subjective, and Fischhoff et al.~\cite{fischhoff1982lay} caution us that patterns in quantitative risk are not neatly explainable by general propensities or by the condescending assumption that lay people are incapable of understanding risk: ``people's perceptions may sometimes be erroneous but they are seldom stupid or irrational.'' While we qualitatively explore outliers in our supplemental material in more depth, we note that even our expert forecasters in the ``Observer'' condition physically stationed on the summit of Mount Washington still disagreed about risk.

\section{Discussion}

Our study provides preliminary evidence that the inclusion of icon-based visual summaries of hazard conditions results in more attention to the risks of outdoor activities. Furthermore, the inclusion of these icons does not seem to result (admittedly, in the context of a study with incentives for participation and attention) in a consistent decrease of use of more detailed forecast information. To some extent this result is to be expected: visual emphasis is a common tool for highlighting pertinent information. However, these hazard icons are simple reiterations and re-expressions of textual information, and so differ in some key respects from other forms of summaries which may focus on more clearly defined analytical tasks like aggregation or filtering~\cite{sarikaya2018design}. Rather, these summaries are intended to have rhetorical force: to underline risks, moderate behavior, and allow the consumers of forecasts to be better prepared, or even change their mind about visiting at all. This rhetoric cannot be patronizing (or it will be ignored or discarded), nor can it obscure the underlying information (to use an example brought up in the workshop, while a casual hiker might be dissuaded by a forecast that contains no information other than a warning that conditions are too dangerous for a visit, the search and rescue teams that use detailed forecast information to plan their rescue events would be hampered, perhaps fatally, by such omissions). That these assessments of risk happen under conditions of high uncertainty, where public expectations often diverge from the statistical realities of weather forecasting~\cite{joslyn2010communicating}, suggests even more caution.

Omnipresent during this study was the notion that ``all visualization research, no matter how superficially apolitical or trivial, has a moral
character''~\cite{correll2019ethical}. Real lives are at stake with this proposed intervention: small changes in wording, moving around crucial information, or the addition of confusing or misinterpreted design elements could result in loss of life or limb for the hundreds of thousands of visitors to Mount Washington every year. Medical ethics often discusses the importance of balancing beneficence (the benefit to the patient) with non-maleficence (the harm to the patient)~\cite{andersson2010no}. While design studies in visualization often assess the \textit{benefits} of new designs (in terms of, say, task performance or accuracy), we believe that proper respect for non-maleficence (which might involve, for instance, avoiding metaphorical ``over-treatment'', or testing against ``placebos''), is an underappreciated component of design study work.

\noindent
\\
\textbf{Acknowledgments:} This research was made possible by Congressionally Directed Spending funding from the National Oceanic and Atmospheric Administration made possible United States Senator Jeanne Shaheen.


\bibliographystyle{eg-alpha-doi} 
\bibliography{mwo}       



\end{document}